\documentclass[
    ,final            
  ]
  {aipproc}

\layoutstyle{6x9}

\begin{document}

\title{Swift UVOT Observations of Core-Collapse SNe}




\author{Peter J. Brown}{
  address={Pennsylvania State University,
                 Dept of Astronomy \& Astrophysics,
                 University Park, PA 16802}
}

\author{Peter W. A. Roming}{
  address={Pennsylvania State University,
                 Dept of Astronomy \& Astrophysics,
                 University Park, PA 16802}
}
\author{Daniel E. Vanden Berk}{
  address={Pennsylvania State University,
                 Dept of Astronomy \& Astrophysics,
                 University Park, PA 16802}
}

\author{Stephen T. Holland}{
  address={Astrophysics Science Division,
		X-Ray Astrophysics Branch, Code 662,
                 NASA Goddard Space Flight Center,
                 Greenbelt, MD 20771},
altaddress={Universities Space Research Association,
		10211 Wincopin Circle, Columbia MD 21044} 
}
\author{Stefan Immler}{
  address={Astrophysics Science Division,
		X-Ray Astrophysics Branch, Code 662,
                 NASA Goddard Space Flight Center,
                 Greenbelt, MD 20771},
altaddress={Universities Space Research Association,
		10211 Wincopin Circle, Columbia MD 21044} 
}
\author{Peter Milne}{
  address={Dept of Astronomy and Steward Observatory, 
		University of Arizona, 
		Tucson, AZ 85721}
}

\begin{abstract}
We review recent UV observations of core-collapse supernovae (SNe) 
with the Swift Ultra-violet/Optical Telescope (UVOT) during its first two years.  
Rest-frame UV photometry is useful for differentiating SN types by exploiting 
the UV-optical spectral shape and more subtle UV features.  This is 
useful for the real-time classification of local and high-redshift SNe 
using only photometry.  Two remarkable SNe Ib/c were observed with UVOT -- 
SN2006jc was a UV bright SN Ib.  Swift observations of GRB060218/SN2006aj 
began shortly after the explosion and show a UV-bright peak followed 
by a UV-faint SN bump.  UV observations are also useful for 
constraining the temperature and ionization structure of SNe IIP.  
Rest-frame UV observations of all types are important for understanding 
the extinction, temperature, and bolometric luminosity of SNe and to 
interpret the observations of high redshift SNe observed at optical wavelengths.

\end{abstract}

\maketitle


\section{Introduction}

In addition to the follow-up observations of Gamma Ray Bursts (GRBs), 
the Swift satellite [1] observes supernovae (SNe) 
discovered primarily from the ground.  These observations capitalize on 
Swift's rapid response capability, flexible short term scheduling, and 
multi-wavelength capabilities, particularly UV observations with the 
Ultra-Violet/Optical Telescope (UVOT; [2] ) and 
X-ray observations with the X-Ray Telescope (XRT; [3]).  
The UV observations of SNe Ia and X-ray observations performed by Swift 
are discussed elsewhere in this volume [4,5].  Here we focus on 
the use of UV photometry to distinguish SN types, and 
describe UVOT observations of several core-collapse SNe.  
A list of the core-collapse SNe observed in the first 
two years of Swift UVOT (2005 and 2006) is given in Table 1.  
\\
\\
\begin{table}[h]
\begin{tabular}{llllll} 
\hline
  \tablehead{1}{l}{b}{Ib} &
  \tablehead{1}{l}{b}{Ib/c} &
  \tablehead{1}{l}{b}{Ic} & 
  \tablehead{1}{l}{b}{IIP} &
  \tablehead{1}{l}{b}{IIb} &
  \tablehead{1}{l}{b}{IIn} \\
\hline
SN2006dn & SN2005bf & SN2005da & SN2005cs & SN2006T & SN2006bv \\
SN2006jc & SN2006lc & SN2005ek & SN2006at &         &          \\
SN2006lt &          & SN2006aj & SN2006bc &         &          \\
         &          &          & SN2006bp &         &          \\
\hline
\end{tabular}
\caption{Core-Collapse SNe observed by Swift in 2005-6.}
\label{tab:b}
\end{table}


\begin{figure}
  \includegraphics[height=.58\textheight]{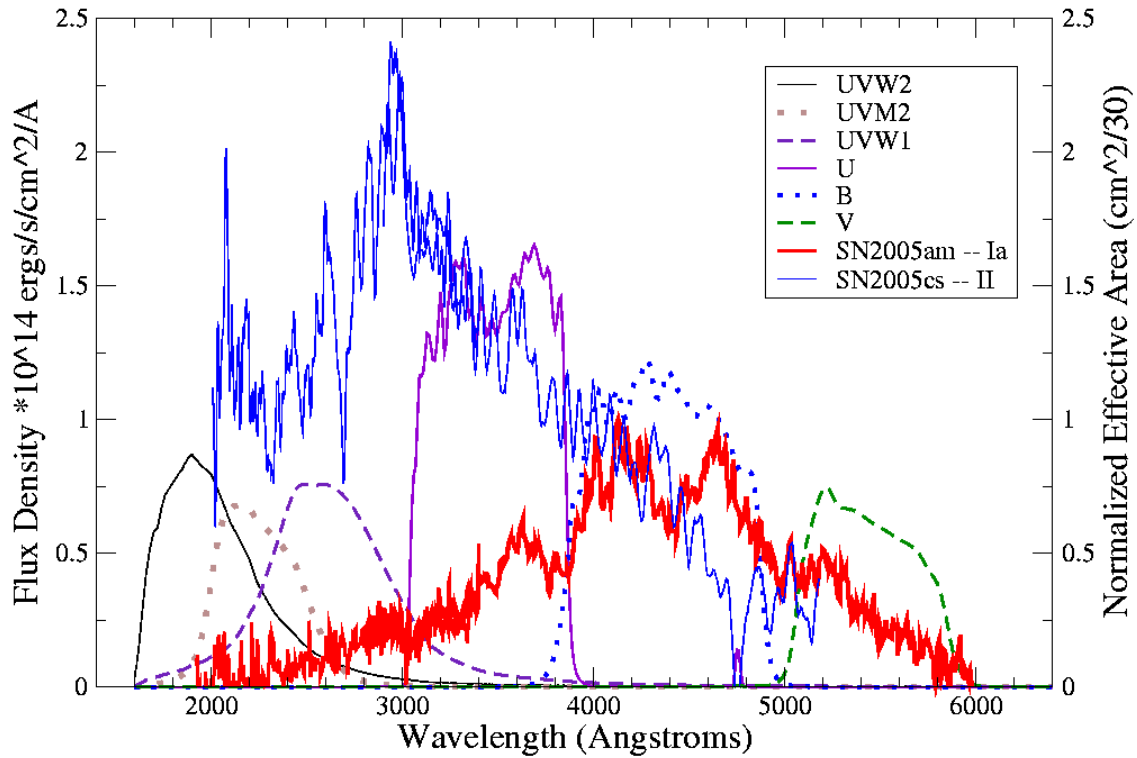}
  \caption{UVOT grism spectra of SNe 2005am (Ia) and 2005cs (II) showing 
the difference in UV-optical shape of the spectral energy distribution [6,7].
The effective area curves of the UVOT filters are also shown (from 
Poole et al. in preparation.)     
}
\end{figure}

\section{SN Phototyping with UV Photometry}

The UV properties of young SNe I and II are very different.  
Young SNe II have extremely high photospheric temperatures, 
resulting in a large quantity of UV photons.  The atmospheres 
of SNe I (including the subtypes Ia, Ib, and Ic) 
are usually much cooler when observed.  This creates a flux 
continuum that is severely depleted in the UV already, with 
many lines from iron peak elements blocking much of the 
remaining UV flux. 
The difference is well represented in Fig 1 which compares UV grism 
spectra obtained with Swift UVOT for the SN Ia 2005am [6] 
and the SN II 2005cs [7].  Also shown are the effective 
area curves of the Swift UVOT filters.  The strong difference between 
the UV-optical flux distribution allows SNe II to be 
distinguished from SNe I without a spectrum [8,9], 
expanding the idea of phototyping which has been explored by 
several authors in the optical bands 
[10,11].  
The UV-optical difference is especially important for classifying high redshift 
SNe observed in the optical [8,12]. 

\clearpage
\begin{figure}
  \includegraphics[height=0.9\textheight]{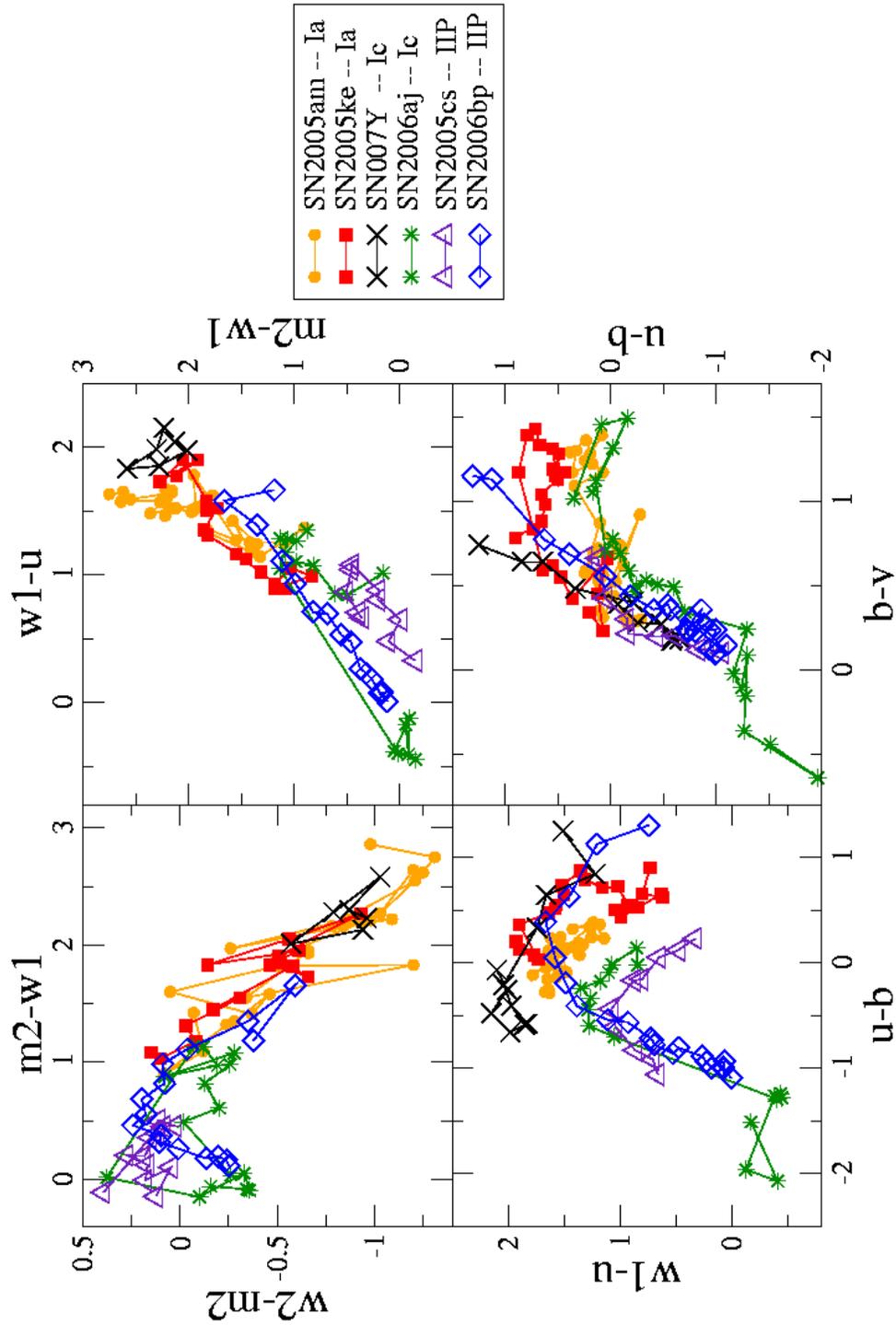}
  \caption{Color-color plots showing the differentiation of different 
SN types using rest-frame UV and optical photometry.  The colors are 
given in the UVOT's Vega system (Poole et al. in preparation), but 
the u, b, and v colors are close to the Johnson system so that 
the lower right panel is comparable to the U-B vs. B-V plots in 
[11].  Extinction vectors in the UV depend on 
the spectral shape of the source and the extinction curve chosen, 
so are not plotted here. }


\end{figure}
\clearpage


UVOT's 6 filters, spanning the UV and optical from 1700 \AA~to 6000 \AA, 
provide excellent coverage of the bluer wavelengths where SNe are most 
clearly differentiated.  Color-color plots using the 6 filters are 
displayed in Fig 2.  They utilize the photometry of two SNe Ia 
(2005am and ke), two SNe Ic (2006aj and 2007Y), 
and two SNe II (2005cs and bp).  One advancement with the UVOT photometry 
is the better time sampling out to later times than the generally sparse UV 
observations available for previous studies.

There is a high degree of overlap in the optical colors, as shown in the 
lower right hand panel of Fig. 2 (see equivalent plots in [11]).  
The general trend is that as more UV information is used, the SNe II 
become better separated from the SNe Ia.  SNe II are in general bluer, 
though SNe Ia are extremely faint in the UVM2 filter [6], resulting 
in the UV colors of SNe Ia being very red in uvm2-uvw1 but blue in uvw2-uvm2.  
In all colors, however, the reddening of the SNII spectrum with time 
makes its colors more similar to a Ia as it becomes older.  
Milky Way extinction will also have a similar 
effect, though a full discussion of that is beyond the scope of this paper.  
Thus a SN II cannot be conclusively distinguished 
from a Ia without addition information on the age and reddening. 
The cadence of the SN search, behavior of light curve, 
or an approximate absolute magnitude might break the degeneracy.

Differentiating SNe Ib/c from SNe Ia is more difficult.  
The UV colors distinguish the Ic SN2006aj, which in the optical 
colors transitions from near the blue SNe II to the red SNe Ia but 
in the UV colors remains closer to the locus of SNe II.  Another Ic SN2007Y, 
however, has a different color evolution that is harder to distinguish from SNe Ia. 
As both SNe Ic are peculiar it is not known what range of UV colors 
SNe Ib/c could have.



\section{SN 2006\lowercase{jc} -- I\lowercase{b}  }
Spectral features from various SN types were identified in early spectra 
of SN2006jc which is perhaps best classified as a peculiar Ib,
though from comparisons with Fig 2, its colors [13]
would have led to it being confidently 
(but erroneously) phototyped as a young SN II.  The blue optical continuum 
is attributed to Fe II emission lines 
blended into a pseudocontinuum [14].  UV grism spectra with UVOT also reveal 
emission from Mg II (Immler et al. 2007, in preparation) 
which suggests circumstellar matter interaction is also a contributor to the 
UV.  X-ray emission was detected with Swift XRT and Chandra [15].  
The blue colors persist and there is little color evolution as the SN fades.  
Light curves of SN2006jc in three filters are displayed in the left panel of Fig. 3.  
The full curves will be presented and discussed by Modjaz et al. (in preparation).

\begin{figure}
  \includegraphics[height=.4\textheight]{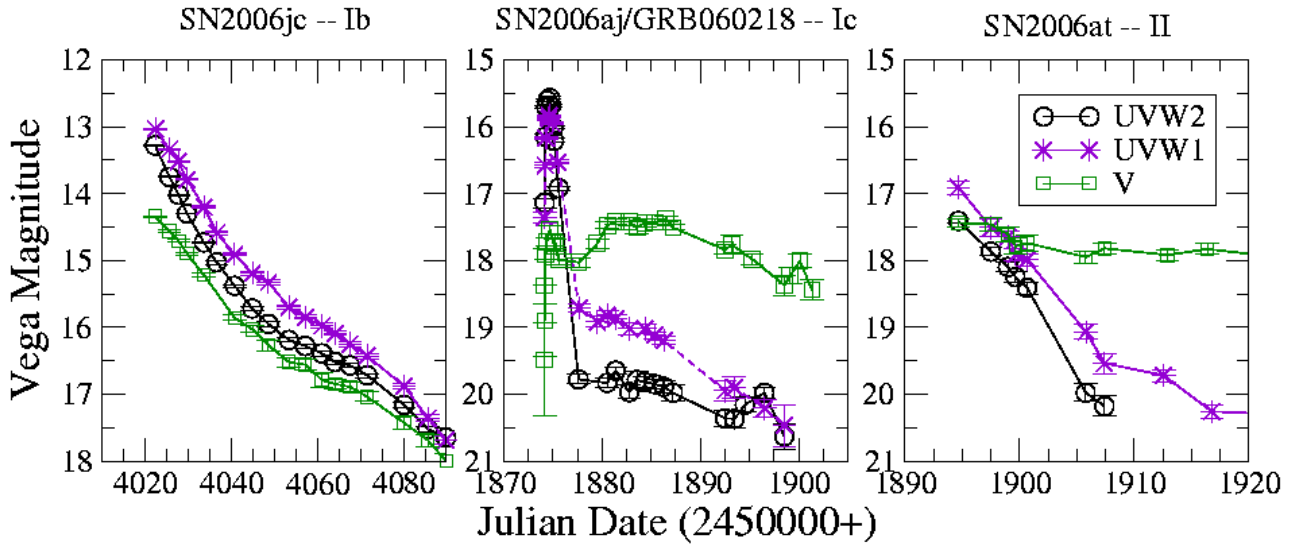}
  \caption{Swift UVOT photometry of SN2006jc, a peculiar Ib (left panel), 
and SN2006aj, a SN Ic associated with GRB060218 (center panel), 
and SN2006at, a SN II (right panel). For clarity, only two UV filters 
(UVW2 and UVW1) are shown along with the V curves for comparison.  }
\end{figure}

\section{GRB060218/SN~2006\lowercase{aj} -- I\lowercase{c}  }

Swift triggered on GRB060218 and was able to observe the accompanying SN 
likely within minutes of core-collapse.  The UVOT light curves exhibit a 
peculiar double peaked structure, as seen in the central panel of Fig 3.  
The first peak is UV-bright, likely 
from the shock breaking out of a wind-blown cocoon [16-18].  Self-absorbed synchrotron 
radiation has also been used to explain it [19].  The first peak is 
followed about ten days later by a second, broader peak driven by
radioactive decay in the SN ejecta.  
The relative brightness of the two peaks is strongly wavelength dependent.  
While of equal brightness in the V band, the shock peak is 4 magnitudes 
brighter than the SN peak in the UVW2 filter.  The detection of 
the SN in the UV is important for estimating the brightness 
of GRB-SNe at higher redshift observed in the optical.

\section{SN\lowercase{e} II}

The strong UV flux at early times makes SNe II attractive targets for 
UV observations.  Three SNe II have been well observed by Swift, 
namely SNe 2005cs, 2006at, and 2006bp.  Current efforts have focused 
on modelling SNe 2005cs and 2006bp which had excellent optical spectroscopic 
coverage to accompany the Swift observations.  Particularly at 
early times, UV observations put strong 
constraints on the line of sight extinction.  As the photosphere 
cools, the UV flux drops dramatically due to the temperature and the 
subsequent line blanketing of iron-peak elements.  Sample lightcurves of 
SN2006at are displayed in the right panel of Fig 3.  The steeper decay at 
shorter wavelengths is a reflection of the rapid reddening of the SED.  
UV observations, which are a probe of the strength of the line blanketing, 
provide constraints on 
the temperature/ionization state of the ejecta.  The observations and 
preliminary modelling for SNe 2005cs and 2006bp have been presented [7,20] 
with more detailed, quantitative results to be presented by 
Dessart et al. (2007 in prep).  



 



\bibliographystyle{aipprocl} 

\bibliography{sample}




\end{document}